\documentclass[12pt,a4paper,english,aps,superscriptaddress,groupedaddress,showpacs,showkeys]{revtex4}

\usepackage[T1]{fontenc}
\usepackage[latin9]{inputenc}
\usepackage{babel}
\usepackage{textcomp}
\usepackage{amsmath}
\usepackage{amssymb}
\usepackage{esint}
\usepackage[unicode=true]
 {hyperref}
\usepackage{breakurl}

\makeatletter

\@ifundefined{textcolor}{}
{%
 \definecolor{BLACK}{gray}{0}
 \definecolor{WHITE}{gray}{1}
 \definecolor{RED}{rgb}{1,0,0}
 \definecolor{GREEN}{rgb}{0,1,0}
 \definecolor{BLUE}{rgb}{0,0,1}
 \definecolor{CYAN}{cmyk}{1,0,0,0}
 \definecolor{MAGENTA}{cmyk}{0,1,0,0}
 \definecolor{YELLOW}{cmyk}{0,0,1,0}
}

\usepackage{dcolumn}
\usepackage{bm}
\usepackage{amsfonts}

\makeatother

\begin{document}

\title{Constructing the Hamiltonian for a free 1D KFGM particle in an interval}

\author{Techapon Kampu}

\homepage{https://orcid.org/0009-0003-1427-8409}

\email{[techaponk65@nu.ac.th]}

\affiliation{The Institute for Fundamental Study (IF), Naresuan University, Phitsanulok
65000, Thailand}

\author{Salvatore De Vincenzo}

\homepage{https://orcid.org/0000-0002-5009-053X}

\email{[salvatored@nu.ac.th]}

\affiliation{The Institute for Fundamental Study (IF), Naresuan University, Phitsanulok
65000, Thailand}

\thanks{Salvatore De Vincenzo would like to dedicate this paper to the memory
of Luigi Mondino, physicist, teacher, tutor, and friend.}

\date{December 27, 2025}

\begin{abstract}
\noindent We analyze the problem of a free 1D Klein-Fock-Gordon-Majorana
(KFGM) particle in an interval. By free, we mean that there is no
potential within the interval and that its walls are penetrable; hence,
the pertinent energy current density does not vanish at the walls.
Certainly, quantization in an interval is not trivial because certain
restrictions imposed by the domains of the operators involved arise.
Here, our objective is to obtain the Hamiltonian for these particles.
In practice, the Feshbach-Villars (FV)--free Hamiltonian is the proper
operator for characterizing them and is a function of the momentum
operator. Additionally, a Majorana condition must also be imposed
on the wavefunctions on which these two operators can act. Thus, we
start by calculating the pseudo self-adjoint momentum operator. A
three-parameter set of boundary conditions (BCs) constitutes its domain.
Up to this point, the domain of the Hamiltonian is induced by the
domain of the momentum operator; however, we ensure that only the
BCs for which the energy current density has the same value at each
end of the interval are in its domain. All these BCs essentially belong
to a one-parameter set of BCs. Moreover, because the FV equation is
invariant under the operation of parity, the parity-transformed wavefunction
is also a solution of this equation, which further restricts the domain
of the free FV Hamiltonian. Finally, knowing the most general three-parameter
set of BCs for the pseudo self-adjoint FV Hamiltonian for a 1D KFGM
particle in an interval, we find that only two BCs can remain within
the domain of the FV--free Hamiltonian: the periodic BC and the antiperiodic
BC. These BCs are satisfied by both the two-component FV wavefunction,
with these components being related, and the one-component KFG wavefunction,
which can be real or imaginary.
\end{abstract}



\maketitle

\section{Introduction}

\noindent As discussed in a few articles and books, for example, in
Refs. \cite{RefA,RefB,RefC,RefD}, the Schr\"odinger Hamiltonian
for a free one-dimensional (1D) Schr\"odinger particle moving in
a penetrable or transparent-walled interval, e.g., $\Omega=[a,b]$,
is a function of its momentum operator squared. As a consequence,
the boundary conditions (BCs) present in its domain are the same as
those in the domain of the self-adjoint momentum operator, namely,
$\Phi_{\mathrm{S}}(b,t)=\mathrm{e}^{\mathrm{i}\,\theta}\,\Phi_{\mathrm{S}}(a,t)$,
with $\theta\in[0,2\pi)$; however, in addition, $(\hat{\mathrm{p}}_{\mathrm{S}}\Phi_{\mathrm{S}})(x,t)$,
i.e., $\Phi'_{\mathrm{S}}(x,t)$, must also satisfy this BC, i.e.,
$\Phi'_{\mathrm{S}}(b,t)=\mathrm{e}^{\mathrm{i}\,\theta}\,\Phi'_{\mathrm{S}}(a,t)$
(the prime symbol is used to denote the first spatial derivative of
a function hereinafter). Thus, we have a one-parameter general set
of BCs for this Hamiltonian, which is also a self-adjoint operator
\cite{RefB}. Clearly, the well-known Dirichlet BC, i.e., $\Phi_{\mathrm{S}}(b,t)=\Phi_{\mathrm{S}}(a,t)=0$,
is not included in this general set of BCs. Indeed, if this BC was
included, the interval could not be considered penetrable but impenetrable;
i.e., the relation $j_{\mathrm{S}}(b,t)=j_{\mathrm{S}}(a,t)=0$, i.e.,
the so-called impenetrability condition at the extremes of the interval,
would be verified, and the particle would not truly be a free particle.
We recall that, as a consequence of the self-adjointness of the Hamiltonian,
the usual probability current density $j_{\mathrm{S}}=j_{\mathrm{S}}(x,t)$
always satisfies the relation $j_{\mathrm{S}}(b,t)=j_{\mathrm{S}}(a,t)$.
Thus, the local observable suitable for characterizing a 1D Schr\"odinger
particle living in a finite interval (in an impenetrable interval
or in an interval with transparent walls) is precisely $j_{\mathrm{S}}$. 

The number of BCs characterizing a free 1D Schr\"odinger particle
in an interval can be further restricted if one takes into account
that the free Schr\"odinger equation has as its solution the wavefunction
$\Phi_{\mathrm{S}}(x,t)$ and its parity-transformed wavefunction
$(\hat{\Pi}_{\mathrm{S}}\Phi_{\mathrm{S}})(x,t)=\Phi_{\mathrm{S}}(a+b-x,t)$.
Thus, this fact leads to only two BCs: the periodic BC (i.e., $\theta=0$),
$\Phi_{\mathrm{S}}(b,t)=\Phi_{\mathrm{S}}(a,t)$ and $\Phi_{\mathrm{S}}'(b,t)=\Phi_{\mathrm{S}}'(a,t)$;
and the antiperiodic BC (i.e., $\theta=\pi$), $\Phi_{\mathrm{S}}(b,t)=-\Phi_{\mathrm{S}}(a,t)$
and $\Phi_{\mathrm{S}}'(b,t)=-\Phi_{\mathrm{S}}'(a,t)$. Naturally,
only for these two BCs is there no distinction between the two walls
of the interval, as is to be expected for a particle moving freely
within an interval (see Appendix A). For example, in Ref. \cite{RefE},
the antiperiodic BC was used in the description of a genuine free
1D Schr\"odinger particle in an interval; however, the domain of
the (free) Hamiltonian operator was not given explicitly. Similarly,
the problem of the free 1D Dirac particle in the finite interval was
considered in Ref. \cite{RefF}. Considering that the momentum operator
under the parity operation must be transformed as a polar vector and
that the Hamiltonian operator function of this operator must be self-adjoint
and that it commutes with the combined operations of charge conjugation,
parity and time reversal (i.e., the free time-dependent Dirac equation
is form invariant under this combined discrete symmetry), the periodic
and antiperiodic BCs were also obtained. Certainly, the usual Dirac
probability current density in these two physical situations does
not vanish at the interval walls. Incidentally, the BCs for a 1D Weyl-Majorana
particle in an interval can be only the periodic and the antiperiodic
BCs; that is, only these two BCs can be imposed on the wavefunctions
that describe this type of particle (with a potential within the interval,
or not) \cite{RefG}.

The nontrivial local observable that appears to be physically acceptable
for completely characterizing a strictly neutral 1D Klein-Fock-Gordon
(KFG) particle \cite{RefH,RefI,RefJ} (or a strictly neutral 1D Feshbach-Villars
(FV) particle \cite{RefK}), i.e., a 1D KFG-Majorana (KFGM) particle
\cite{RefL,RefM} (or a 1D FV-Majorana (FVM) particle) that lies within
the interval $\Omega=[a,b]$ (confined to the interval or restricted
in the interval with transparent walls), is given by
\begin{equation}
j_{\mathrm{en}}=j_{\mathrm{en}}(x,t)=-\frac{\hbar^{2}}{2\mathrm{m}}\,\frac{1}{2}\left[\,\left((\hat{\tau}_{3}+\mathrm{i}\hat{\tau}_{2})\Phi'\right)^{\dagger}(\hat{\tau}_{3}+\mathrm{i}\hat{\tau}_{2})\dot{\Phi}-\left((\hat{\tau}_{3}+\mathrm{i}\hat{\tau}_{2})\Phi\right)^{\dagger}(\hat{\tau}_{3}+\mathrm{i}\hat{\tau}_{2})\dot{\Phi}'\,\right],
\end{equation}
where $\Phi=\Phi(x,t)$ is a two-component FV wavefunction \cite{RefN},
$\dot{\Phi}=\partial\Phi/\partial t$ hereinafter, and $\hat{\tau}_{2}=\hat{\sigma}_{y}$
and $\hat{\tau}_{3}=\hat{\sigma}_{z}$ are Pauli matrices. We recall
that the KFG and FV wave equations are completely equivalent, as long
as the mass of the particle is not zero. The energy current density
in Eq. (1) is a real quantity when it is calculated for a state $\Phi$
satisfying a Majorana condition, which is physically adequate \cite{RefN}.
Additionally, as a consequence of the pseudo self-adjointness (i.e.,
generalized self-adjointness) of the Hamiltonian operator of the system,
it follows that
\begin{equation}
j_{\mathrm{\mathrm{en}}}(b,t)=j_{\mathrm{\mathrm{en}}}(a,t).
\end{equation}
A necessary condition for having a confined 1D KFGM particle to an
interval is that the energy current density $j_{\mathrm{en}}$ (i.e.,
the flow of energy) disappears at the walls of that interval, i.e.,
$j_{\mathrm{\mathrm{en}}}(b,t)=j_{\mathrm{\mathrm{en}}}(a,t)=0$.
On the other hand, a necessary condition for having a free 1D KFGM
particle in an interval is that $j_{\mathrm{en}}$ simply obeys the
relation given in Eq. (2), i.e., $j_{\mathrm{\mathrm{en}}}(b,t)=j_{\mathrm{\mathrm{en}}}(a,t)$.
In the latter case, we say that the interval has transparent walls
(or walls that allow leaks) \cite{RefN}. Let us also note that we
have not used here the usual KFG current density $j=j(x,t)$ (this
density is $\mathrm{i}\hbar/4\mathrm{m}$ multiplied by the term in
brackets in Eq. (1), but without any of the time derivatives). The
reason is that this local observable vanishes everywhere whenever
a Majorana condition is imposed on $\Phi$. Therefore, $j$ does not
allow distinguishing between penetrable and impenetrable BCs if the
particle is a 1D KFGM particle \cite{RefN} (see Appendix B).

The main objective of our paper is to obtain the pseudo self-adjoint
Hamiltonian operator that is physically adequate for characterizing
a truly free 1D KFGM particle in an interval, i.e., its specific domain
(which includes the proper BCs). This operator, the FV--free Hamiltonian,
is a function of the momentum operator for this particle; hence, the
latter operator plays an important role in the construction of the
Hamiltonian. Additionally, a Majorana condition must also be imposed
on the two-component wavefunctions on which these two operators can
act. Specifically, in the next section, we obtain the domain of the
pseudo self-adjoint momentum operator that is consistent with the
Majorana condition. We find that a three-parameter general family
of BCs constitutes the domain of this operator. In Section III, we
show how the domain of the free FV Hamiltonian is induced by the domain
of the momentum operator. After that, we first ensure that, in the
domain of the free FV Hamiltonian, only the BCs that satisfy the relation
in Eq. (2) remain. Because the (free) FV equation is invariant under
parity (i.e., space reflection), the parity-transformed wavefunction
is also a solution of this equation, which further restricts the domain
of the Hamiltonian. To determine whether the BCs that emerge here
(written in terms of the two-component FV wavefunction) can actually
be in the domain of the pseudo self-adjoint--free Hamiltonian operator,
we use the relation that connects the FV wavefunctions with the one-component
KFG wavefunctions and transforms these BCs into BCs for the KFG wavefunction.
The reason for this is that we already know the most general three-parameter
set of BCs written in terms of the one-component KFG wavefunction
for the pseudo self-adjoint FV Hamiltonian that describes a 1D KFGM
particle in an interval. Thus, only BCs that are in this general three-parameter
set can be part of the domain of the free FV Hamiltonian operator,
which is also a pseudo self-adjoint operator. In the end, only two
quantum BCs remain in the domain of the free FV Hamiltonian: the periodic
BC and the antiperiodic BC. Finally, we discuss our results in Section
IV and complement what was mentioned in the second paragraph of this
Introduction in Appendix A. Similarly, in Appendix B, we provide details
that complement the ideas presented in the third paragraph of this
Introduction, whereas a particular ``extra'' set of BCs that arises
in Section III is analyzed in Appendix C. 

\section{The momentum operator for a 1D KFGM particle }

\noindent The Hamiltonian operator describing a free 1D KFGM particle,
i.e., a free 1D KFG particle that is also strictly neutral, is the
free FV Hamiltonian, namely, 
\begin{equation}
\hat{\mathrm{h}}=(\hat{\tau}_{3}+\mathrm{i}\hat{\tau}_{2})\frac{\hat{\mathrm{p}}^{2}}{2\mathrm{m}}+\mathrm{m}c^{2}\hat{\tau}_{3}=(\hat{\tau}_{3}+\mathrm{i}\hat{\tau}_{2})\frac{1}{2\mathrm{m}}\hat{\mathrm{p}}\hat{\mathrm{p}}+\mathrm{m}c^{2}\hat{\tau}_{3},
\end{equation}
where
\begin{equation}
\hat{\mathrm{p}}=-\mathrm{i}\hbar\,\hat{1}_{2}\,\frac{\partial}{\partial x}
\end{equation}
is the momentum operator, $\hat{\tau}_{2}=\hat{\sigma}_{y}$, $\hat{\tau}_{3}=\hat{\sigma}_{z}$
and $\hat{1}_{2}$ is the $2\times2$ identity matrix \cite{RefO,RefP}.
Moreover, the functions on which $\hat{\mathrm{p}}$ (and $\hat{\mathrm{h}}$)
act are two-component wavefunctions of the form $\Phi=\Phi(x,t)=\left[\,\phi_{1}\;\,\phi_{2}\,\right]^{\mathrm{T}}=\left[\,\phi_{1}(x,t)\;\,\phi_{2}(x,t)\,\right]^{\mathrm{T}}$
(the symbol $^{\mathrm{T}}$ represents the transpose of a matrix)
and must satisfy a Majorana condition, i.e.,  $\Phi=\Phi_{c}\equiv\hat{\tau}_{1}\Phi^{*}$
($\Rightarrow\phi_{2}=\phi_{1}^{*}$) or $\Phi=-\Phi_{c}=-\hat{\tau}_{1}\Phi^{*}$
($\Rightarrow\phi_{2}=-\phi_{1}^{*}$), where $\Phi_{c}$ is the charge-conjugate
wavefunction, the asterisk denotes the complex conjugation operation
and $\hat{\tau}_{1}=\hat{\sigma}_{x}$ is a Pauli matrix. Clearly,
the two components of $\Phi$ are not independent \cite{RefM}.

The formal generalized adjoint of $\hat{\mathrm{p}}$ is defined as
follows: 
\begin{equation}
\hat{\mathrm{p}}_{\mathrm{adj}}\equiv\hat{\tau}_{3}\,\hat{\mathrm{p}}^{\dagger}\,\hat{\tau}_{3}=-\mathrm{i}\hbar\,\hat{1}_{2}\,\frac{\partial}{\partial x},
\end{equation}
where $\hat{\mathrm{p}}^{\dagger}$ is the usual formal Hermitian
conjugate of $\hat{\mathrm{p}}$. Clearly, the actions of $\hat{\mathrm{p}}_{\mathrm{adj}}$
and $\hat{\mathrm{p}}$ are equal. The definition given in Eq. (5)
leads to the following relation: 
\begin{equation}
\langle\langle\hat{\mathrm{p}}_{\mathrm{adj}}\Psi,\Phi\rangle\rangle=\langle\langle\Psi,\hat{\mathrm{p}}\Phi\rangle\rangle,
\end{equation}
where the indefinite (or improper) inner product is defined as follows:
\begin{equation}
\langle\langle\Psi,\Phi\rangle\rangle\equiv\int_{\Omega}\mathrm{d}x\,\Psi^{\dagger}\hat{\tau}_{3}\Phi=\langle\langle\Phi,\Psi\rangle\rangle^{*}
\end{equation}
($\Psi=\left[\,\psi_{1}\;\,\psi_{2}\,\right]^{\mathrm{T}}$ and $\Phi=\left[\,\phi_{1}\;\,\phi_{2}\,\right]^{\mathrm{T}}$).
We assume that the particle lives in the interval $\Omega=[a,b]$.
The inner product in Eq. (7) is defined over an indefinite inner product
space; in our case, a $\hat{\tau}_{3}$-space \cite{RefQ,RefR}. The
relation in Eq. (6) defines the generalized adjoint of $\hat{\mathrm{p}}$
on the $\hat{\tau}_{3}$-space, i.e., $\hat{\mathrm{p}}_{\mathrm{adj}}$,
where $\Psi\in\mathrm{\mathcal{D}}(\hat{\mathrm{p}}_{\mathrm{adj}})$,
the domain of $\hat{\mathrm{p}}_{\mathrm{adj}}$, and $\Phi\in\mathrm{\mathcal{D}}(\hat{\mathrm{p}})$,
the domain of $\hat{\mathrm{p}}$. These domains are linear subsets
of the $\hat{\tau}_{3}$-space on which we allow the operators $\hat{\mathrm{p}}$
and $\hat{\mathrm{p}}_{\mathrm{adj}}$ to act. Certainly, BCs are
a key part of these domains, and $\mathrm{\mathcal{D}}(\hat{\mathrm{p}})$
and $\mathrm{\mathcal{D}}(\hat{\mathrm{p}}_{\mathrm{adj}})$ do not
have to be the same.

If $\hat{\mathrm{p}}$ satisfies the relation 
\begin{equation}
\langle\langle\hat{\mathrm{p}}\Psi,\Phi\rangle\rangle=\langle\langle\Psi,\hat{\mathrm{p}}\Phi\rangle\rangle,
\end{equation}
the operator $\hat{\mathrm{p}}$ is a generalized self-adjoint operator
or a pseudo self-adjoint operator, i.e., $\hat{\mathrm{p}}=\hat{\mathrm{p}}_{\mathrm{adj}}$;
hence, the actions of $\hat{\mathrm{p}}$ and $\hat{\mathrm{p}}_{\mathrm{adj}}$
must be equal (in fact, they are equal), and their domains must also
be the same. Specifically, by virtue of one integration by parts,
$\hat{\mathrm{p}}$ and $\hat{\mathrm{p}}_{\mathrm{adj}}$ satisfy
the following relation: 
\begin{equation}
\langle\langle\hat{\mathrm{p}}_{\mathrm{adj}}\Psi,\Phi\rangle\rangle=\langle\langle\Psi,\hat{\mathrm{p}}\Phi\rangle\rangle+f[\Psi,\Phi;\Omega],
\end{equation}
where the term evaluated at the boundaries of $\Omega$ is given by
\begin{equation}
f[\Psi,\Phi;\Omega]=\mathrm{i}\hbar\left.\left[\,\Psi^{\dagger}\hat{\tau}_{3}\Phi\,\right]\right|_{a}^{b}=\mathrm{i}\hbar\left\{ \left[\begin{array}{c}
\psi_{1}(b,t)\\
\psi_{2}(a,t)
\end{array}\right]^{\dagger}\left[\begin{array}{c}
\phi_{1}(b,t)\\
\phi_{2}(a,t)
\end{array}\right]-\left[\begin{array}{c}
\psi_{2}(b,t)\\
\psi_{1}(a,t)
\end{array}\right]^{\dagger}\left[\begin{array}{c}
\phi_{2}(b,t)\\
\phi_{1}(a,t)
\end{array}\right]\right\} 
\end{equation}
(in the latter expression, we use the definition $\left.\left[\, g\,\right]\right|_{a}^{b}\equiv g(b,t)-g(a,t)$).
Let us suppose initially that any wavefunction $\Phi\in\mathrm{\mathcal{D}}(\hat{\mathrm{p}})$
satisfies the BC $\Phi(a,t)=\Phi(b,t)=0$, that is, $\phi_{1}(a,t)=\phi_{2}(a,t)=\phi_{1}(b,t)=\phi_{2}(b,t)=0$.
In this case, the boundary term $f[\Psi,\Phi;\Omega]$ in Eq. (10)
vanishes, and Eq. (6) is verified without requiring any BC on the
functions $\Psi\in\mathrm{\mathcal{D}}(\hat{\mathrm{p}}_{\mathrm{adj}})$.
Thus, by initially choosing this BC, one has that $\mathrm{\mathcal{D}}(\hat{\mathrm{p}})\neq\mathrm{\mathcal{D}}(\hat{\mathrm{p}}_{\mathrm{adj}})$
(in fact, $\mathrm{\mathcal{D}}(\hat{\mathrm{p}})\subset\mathrm{\mathcal{D}}(\hat{\mathrm{p}}_{\mathrm{adj}})$);
hence, $\hat{\mathrm{p}}$ is not a pseudo self-adjoint operator. 

If the operator $\hat{\mathrm{p}}$ aims to be a pseudo self-adjoint
operator, i.e., $\hat{\mathrm{p}}=\hat{\mathrm{p}}_{\mathrm{adj}}$,
then $\mathrm{\mathcal{D}}(\hat{\mathrm{p}})=\mathrm{\mathcal{D}}(\hat{\mathrm{p}}_{\mathrm{adj}})$.
Thus, we must allow every wavefunction $\Phi$ inside $\mathrm{\mathcal{D}}(\hat{\mathrm{p}})$
to satisfy some BCs that are less restrictive than the one used above.
Let us propose the following general set of BCs that will be part
of the domain $\mathrm{\mathcal{D}}(\hat{\mathrm{p}})$:
\begin{equation}
\left[\begin{array}{c}
\phi_{1}(b,t)\\
\phi_{2}(a,t)
\end{array}\right]=\hat{\mathrm{N}}\left[\begin{array}{c}
\phi_{2}(b,t)\\
\phi_{1}(a,t)
\end{array}\right],
\end{equation}
where $\hat{\mathrm{N}}$ is an arbitrary complex matrix. Substituting
this relation into $f[\Psi,\Phi;\Omega]$ given in Eq. (10), we obtain
the following expression: 
\begin{equation}
f[\Psi,\Phi;\Omega]=\mathrm{i}\hbar\left\{ \left(\left[\begin{array}{c}
\psi_{1}(b,t)\\
\psi_{2}(a,t)
\end{array}\right]^{\dagger}\hat{\mathrm{N}}-\left[\begin{array}{c}
\psi_{2}(b,t)\\
\psi_{1}(a,t)
\end{array}\right]^{\dagger}\right)\left[\begin{array}{c}
\phi_{2}(b,t)\\
\phi_{1}(a,t)
\end{array}\right]\right\} ,
\end{equation}
and because $f[\Psi,\Phi;\Omega]$ must be zero, and we do not wish
to use the initially chosen BC (because, for the latter, we have that
$\hat{\mathrm{p}}\neq\hat{\mathrm{p}}_{\mathrm{adj}}$), i.e., $\phi_{2}(b,t)=\phi_{1}(a,t)=0$
and $\phi_{1}(b,t)=\phi_{2}(a,t)=0$, we have that 
\begin{equation}
\left[\begin{array}{c}
\psi_{1}(b,t)\\
\psi_{2}(a,t)
\end{array}\right]^{\dagger}\hat{\mathrm{N}}=\left[\begin{array}{c}
\psi_{2}(b,t)\\
\psi_{1}(a,t)
\end{array}\right]^{\dagger}\quad\Rightarrow\quad\left[\begin{array}{c}
\psi_{2}(b,t)\\
\psi_{1}(a,t)
\end{array}\right]=\hat{\mathrm{N}}^{\dagger}\left[\begin{array}{c}
\psi_{1}(b,t)\\
\psi_{2}(a,t)
\end{array}\right].
\end{equation}
Because any wavefunction $\Psi\in\mathrm{\mathcal{D}}(\hat{\mathrm{p}}_{\mathrm{adj}})$
would have to satisfy the same BCs that $\Phi\in\mathrm{\mathcal{D}}(\hat{\mathrm{p}})$
satisfies, i.e., in addition to the BCs in Eq. (13), $\Psi$ should
also satisfy the BCs that are in Eq. (11), namely, 
\begin{equation}
\left[\begin{array}{c}
\psi_{1}(b,t)\\
\psi_{2}(a,t)
\end{array}\right]=\hat{\mathrm{N}}\left[\begin{array}{c}
\psi_{2}(b,t)\\
\psi_{1}(a,t)
\end{array}\right].
\end{equation}
Now, substituting this result into the result given in Eq. (13), we
obtain that $\hat{\mathrm{N}}$ is a unitary matrix, i.e.,
\begin{equation}
\hat{\mathrm{N}}^{\dagger}=\hat{\mathrm{N}}^{-1}.
\end{equation}
This is the condition on $\hat{\mathrm{N}}$ that ensures that any
BC taken from the set given in Eq. (11), which is included in the
domain of $\hat{\mathrm{p}}$, is also in the domain of $\hat{\mathrm{p}}_{\mathrm{adj}}$.
In this way, we make these two domains equal and ensure that $\hat{\mathrm{p}}=\hat{\mathrm{p}}_{\mathrm{adj}}$,
i.e., that $\hat{\mathrm{p}}$ is a pseudo self-adjoint operator.
Because $\hat{\mathrm{N}}$ is a unitary matrix, the result in Eq.
(14) is a four-parameter general family of BCs. 

Let us also note that the Dirac Hamiltonian operator in (1+1) dimensions
$\hat{\mathrm{h}}_{\mathrm{D}}=c\hat{\alpha}\hat{\mathrm{p}}$ ($\hat{\mathrm{p}}=-\mathrm{i}\hbar\hat{1}_{2}\,\partial/\partial x$),
where $x\in\Omega=[a,b]$, in the Weyl representation ($\hat{\alpha}=\hat{\sigma}_{z}$),
is a self-adjoint operator when its domain contains the general set
of BCs given in Eq. (11), with $\hat{\mathrm{N}}$ being a unitary
matrix. In this case, the operator $\hat{\mathrm{h}}_{\mathrm{D}}$
acts on two-component wavefunctions, and the scalar product for these
wavefunctions is the usual one, i.e., $\langle\Phi_{\mathrm{D}},\Psi_{\mathrm{D}}\rangle\equiv\int_{\Omega}\mathrm{d}x\,\Phi_{\mathrm{D}}^{\dagger}\Psi_{\mathrm{D}}$
(see, for example, Ref. \cite{RefS} and references therein).

If we impose $\Psi=\Phi$ in Eq. (8) and in Eq. (9) with the result
$\hat{\mathrm{p}}=\hat{\mathrm{p}}_{\mathrm{adj}}$, we obtain the
following condition: 
\begin{equation}
f[\Phi,\Phi;\Omega]=\mathrm{i}\hbar\left.\left[\,\varrho\,\right]\right|_{a}^{b}=0\quad\Rightarrow\quad\varrho(b,t)=\varrho(a,t)\,,
\end{equation}
where the usual KFG density is given by 
\begin{equation}
\varrho=\varrho(x,t)=\Phi^{\dagger}\hat{\tau}_{3}\Phi=|\phi_{1}|^{2}-|\phi_{2}|^{2}.
\end{equation}
Similar to the relation given in Eq. (9), the following relation between
$\hat{\mathrm{p}}$ and $\hat{\mathrm{p}}_{\mathrm{adj}}$ can also
be written:
\begin{equation}
\langle\langle\hat{\mathrm{p}}_{\mathrm{adj}}\Psi,\hat{\mathrm{E}}\Phi\rangle\rangle=\langle\langle\Psi,\hat{\mathrm{p}}\hat{\mathrm{E}}\Phi\rangle\rangle+f[\Psi,\hat{\mathrm{E}}\Phi;\Omega],
\end{equation}
where $\hat{\mathrm{E}}=\mathrm{i}\hbar\,\hat{1}_{2}\,\partial/\partial t$
is the energy operator. Naturally, if $\Phi\in\mathrm{\mathcal{D}}(\hat{\mathrm{p}})$,
then, automatically, $\hat{\mathrm{E}}\Phi\in\mathrm{\mathcal{D}}(\hat{\mathrm{p}})$.
If $\Phi$ satisfies a BC, then $\hat{\mathrm{E}}\Phi$ satisfies
that same BC. Consequently, any BC included in the general set of
BCs, which we know leads to the cancellation of the boundary term
in Eq. (10), must also cancel out the boundary term in Eq. (18) (compare
the boundary terms in Eqs. (10) and (18) and note that if $\Phi$
satisfies a BC, then $\hat{\mathrm{E}}\Phi$ also satisfy that BC).
We recall that for any of the BCs that are within the general family
of BCs, $\hat{\mathrm{p}}$ is a pseudo self-adjoint operator, i.e.,
$\hat{\mathrm{p}}=\hat{\mathrm{p}}_{\mathrm{adj}}$ (i.e., $\hat{\mathrm{p}}$
acts the same as $\hat{\mathrm{p}}_{\mathrm{adj}}$ and $\mathrm{\mathcal{D}}(\hat{\mathrm{p}})=\mathrm{\mathcal{D}}(\hat{\mathrm{p}}_{\mathrm{adj}})$).
Similarly, if we impose $\Psi=\Phi$ in Eq. (18) with the result $\hat{\mathrm{p}}=\hat{\mathrm{p}}_{\mathrm{adj}}$,
we obtain the following result:
\begin{equation}
f[\Phi,\hat{\mathrm{E}}\Phi;\Omega]=\mathrm{i}\hbar\left.\left[\,\varrho_{\mathrm{en}}\,\right]\right|_{a}^{b}=0\quad\Rightarrow\quad\varrho_{\mathrm{en}}(b,t)=\varrho_{\mathrm{en}}(a,t)\,,
\end{equation}
where the (complex) KFG energy density is given by 
\begin{equation}
\varrho_{\mathrm{en}}=\varrho_{\mathrm{en}}(x,t)=\Phi^{\dagger}\hat{\tau}_{3}\hat{\mathrm{E}}\Phi=\mathrm{i}\hbar\,(\phi_{1}^{*}\,\dot{\phi}_{1}-\phi_{2}^{*}\,\dot{\phi}_{2}),
\end{equation}
where $\dot{\phi}_{1}=\partial\phi_{1}/\partial t$, etc., as usual
\cite{RefP}. The results given in Eqs. (16) and (19) are straightforward
consequences of the pseudo self-adjointness of $\hat{\mathrm{p}}$.
In addition, $j_{\mathrm{en}}$ in Eq. (1) and $\varrho_{\mathrm{en}}$
in Eq. (20) satisfy a continuity equation (this situation is valid
on-shell, i.e., for solutions $\Phi$ of the FV equation. See Appendix
B). 

If we impose any of the two Majorana conditions on the four-parameter
general set of BCs in Eq. (11), i.e., $\phi_{2}=\phi_{1}^{*}\Rightarrow\phi_{1}=\phi_{2}^{*}$,
or $\phi_{2}=-\phi_{1}^{*}\Rightarrow\phi_{1}=-\phi_{2}^{*}$ (with
the condition given in Eq. (15)), then it becomes a three-parameter
general set of BCs. This is because the unitary matrix $\hat{\mathrm{N}}$
must also be a symmetric matrix, namely,
\begin{equation}
\hat{\mathrm{N}}=\hat{\mathrm{N}}^{\mathrm{T}}.
\end{equation}
Similarly, if a Majorana condition is imposed on the KFG density in
Eq. (17), then it automatically vanishes, i.e., $\varrho=0$. Moreover,
if a Majorana condition is imposed on the (complex) KFG energy density
in Eq. (20), it becomes a real quantity, i.e., $\varrho_{\mathrm{en}}=\varrho_{\mathrm{en}}^{*}$.
Certainly, the general set of BCs given in Eq. (11), with $\hat{\mathrm{N}}$
being a unitary and symmetric matrix, can be written in terms of only
one of the components of $\Phi$ ($\phi_{1}$ or $\phi_{2}$) and
the other component can be obtained algebraically ($\phi_{2}=\phi_{1}^{*}$
or $\phi_{1}=\phi_{2}^{*}$, when $\Phi=\Phi_{c}$; and $\phi_{2}=-\phi_{1}^{*}$
or $\phi_{1}=-\phi_{2}^{*}$, when $\Phi=-\Phi_{c}$). All of these
are general sets of BCs that make up the domain of the momentum operator
$\mathrm{\mathcal{D}}(\hat{\mathrm{p}})$ for a 1D KFGM particle.

\section{The Hamiltonian for a free 1D KFGM particle }

\noindent The formal action of the Hamiltonian operator for a free
1D KFGM particle is given in Eq. (3), and its domain is essentially
given by 
\begin{equation}
\mathrm{\mathcal{D}}(\hat{\mathrm{h}})=\mathrm{\mathcal{D}}(\hat{\mathrm{p}}^{2})=\mathrm{\mathcal{D}}(\hat{\mathrm{p}}\hat{\mathrm{p}})\sim\left\{ \,\Phi\mid\Phi\in\mathrm{\mathcal{D}}(\hat{\mathrm{p}})\;\mathrm{and}\;\hat{\mathrm{p}}\Phi\in\mathrm{\mathcal{D}}(\hat{\mathrm{p}})\,\right\} \subset\mathrm{\mathcal{D}}(\hat{\mathrm{p}}).
\end{equation}
Thus, the wavefunctions $\Phi$ belonging to $\mathrm{\mathcal{D}}(\hat{\mathrm{h}})$
must satisfy the same BCs that $\Phi\in\mathrm{\mathcal{D}}(\hat{\mathrm{p}})$
satisfy, i.e., must satisfy any BC included in the set of BCs given
in Eq. (11), with the matrix $\hat{\mathrm{N}}$ satisfying Eqs. (15)
and (21). Additionally, because $\hat{\mathrm{p}}\Phi\in\mathrm{\mathcal{D}}(\hat{\mathrm{p}})$,
the first spatial derivative of $\Phi$, i.e., $\phi_{1}'$ and $\phi_{2}'$,
must also satisfy the set of BCs in Eq. (11), with $\hat{\mathrm{N}}$
fulfilling $\hat{\mathrm{N}}^{\dagger}=\hat{\mathrm{N}}^{-1}$ and
$\hat{\mathrm{N}}=\hat{\mathrm{N}}^{\mathrm{T}}$. The $2\times2$
unitary and symmetric matrix $\hat{\mathrm{N}}$ depends on three
real parameters and can be written as follows:
\begin{equation}
\hat{\mathrm{N}}=\mathrm{e}^{\mathrm{i}\,\mu}\left[\begin{array}{cc}
\mathrm{m}_{0}-\mathrm{i}\,\mathrm{m}_{3} & -\mathrm{i}\,\mathrm{m}_{1}\\
-\mathrm{i}\,\mathrm{m}_{1} & \mathrm{m}_{0}+\mathrm{i}\,\mathrm{m}_{3}
\end{array}\right],
\end{equation}
where the real quantities $\mathrm{m}_{0}$, $\mathrm{m}_{1}$ and
$\mathrm{m}_{3}$ satisfy $(\mathrm{m}_{0})^{2}+(\mathrm{m}_{1})^{2}+(\mathrm{m}_{3})^{2}=1$
(in this article, no parameter is labeled with the symbol $\mathrm{m}_{2}$),
and $\mu\in[0,\pi)$ ($\mathrm{det}(\hat{\mathrm{N}})=\mathrm{e}^{\mathrm{i}\,2\mu}$).
Then, the wavefunctions $\Phi$ on which the Hamiltonian given in
Eq. (3) can act, i.e., the wavefunctions $\Phi$ that make up the
domain of $\hat{\mathrm{h}}$, $\mathrm{\mathcal{D}}(\hat{\mathrm{h}})$,
can only satisfy BCs that are included in the following family of
BCs:
\begin{equation}
\left[\begin{array}{c}
\phi_{1}(b,t)\\
\phi_{2}(a,t)
\end{array}\right]=\hat{\mathrm{N}}\left[\begin{array}{c}
\phi_{2}(b,t)\\
\phi_{1}(a,t)
\end{array}\right]\quad\mathrm{and}\quad\left[\begin{array}{c}
\phi_{1}'(b,t)\\
\phi_{2}'(a,t)
\end{array}\right]=\hat{\mathrm{N}}\left[\begin{array}{c}
\phi_{2}'(b,t)\\
\phi_{1}'(a,t)
\end{array}\right].
\end{equation}
To this point, the domain of $\hat{\mathrm{h}}$ is induced entirely
by the domain of $\hat{\mathrm{p}}$.

However, the Hamiltonian operator with the latter sets of BCs within
its domain is not necessarily a pseudo self-adjoint operator. For
example, the relation given in Eq. (2) is not necessarily verified.
Certainly, if the latter relation holds, the Hamiltonian is not necessarily
a pseudo self-adjoint operator; however, the equality $j_{\mathrm{en}}(b,t)=j_{\mathrm{en}}(a,t)$
is a consequence of the pseudo self-adjointness of $\hat{\mathrm{h}}$.
Thus, we can restrict the set of BCs in Eq. (24) by requiring the
validity of that equality. To obtain the mathematical conditions that
ensure that $j_{\mathrm{en}}(b,t)=j_{\mathrm{en}}(a,t)$, it is convenient
to first rewrite the family of BCs given in Eq. (24) as follows:
\begin{equation}
\left[\begin{array}{c}
\phi_{1}(b,t)\\
\phi_{2}(b,t)
\end{array}\right]=\hat{\mathrm{V}}\left[\begin{array}{c}
\phi_{1}(a,t)\\
\phi_{2}(a,t)
\end{array}\right]\quad\mathrm{and}\quad\left[\begin{array}{c}
\phi_{1}'(b,t)\\
\phi_{2}'(b,t)
\end{array}\right]=\hat{\mathrm{V}}\left[\begin{array}{c}
\phi_{1}'(a,t)\\
\phi_{2}'(a,t)
\end{array}\right],
\end{equation}
where 
\begin{equation}
\hat{\mathrm{V}}=\frac{\mathrm{i}}{\mathrm{m}_{1}}\left[\begin{array}{cc}
-\mathrm{e}^{\mathrm{i}\,\mu} & \mathrm{m}_{0}-\mathrm{i}\,\mathrm{m}_{3}\\
-(\mathrm{m}_{0}+\mathrm{i}\,\mathrm{m}_{3}) & \mathrm{e}^{-\mathrm{i}\,\mu}
\end{array}\right]
\end{equation}
and $\mathrm{m}_{1}\neq0$ ($\mathrm{det}(\hat{\mathrm{V}})=1$).
Certainly, simple algebraic manipulations of the relations given in
Eq. (24) lead to the results in Eqs. (25) and (26). On the other hand,
the case $\mathrm{m}_{1}=0$ is studied in Appendix C. 

Then, evaluating the energy current density given in Eq. (1) at $x=b$,
substituting into there the two relations given in Eq. (25) and their
time derivatives $\dot{\Phi}(b,t)=\hat{\mathrm{V}}\dot{\Phi}(a,t)$
and $\dot{\Phi}'(b,t)=\hat{\mathrm{V}}\dot{\Phi}'(a,t)$, and finally
equaling with $j_{\mathrm{en}}(a,t)$ (see Eq. (2)), we obtain the
result
\begin{equation}
\hat{\mathrm{V}}^{\dagger}(\hat{\tau}_{3}+\mathrm{i}\hat{\tau}_{2})^{\dagger}(\hat{\tau}_{3}+\mathrm{i}\hat{\tau}_{2})\hat{\mathrm{V}}=(\hat{\tau}_{3}+\mathrm{i}\hat{\tau}_{2})^{\dagger}(\hat{\tau}_{3}+\mathrm{i}\hat{\tau}_{2}).
\end{equation}
By substituting the matrix $\hat{\mathrm{V}}$ in Eq. (26) into Eq.
(27), the following three equations are obtained: 
\begin{equation}
1+\mathrm{m}_{0}\cos(\mu)+\mathrm{m}_{3}\sin(\mu)=(\mathrm{m}_{1})^{2},
\end{equation}
\begin{equation}
(\mathrm{e}^{\mathrm{i}\,\mu}+\mathrm{m}_{0}+\mathrm{i}\,\mathrm{m}_{3})^{2}=-(\mathrm{m}_{1})^{2},
\end{equation}
and 
\begin{equation}
(\mathrm{e}^{-\mathrm{i}\,\mu}+\mathrm{m}_{0}-\mathrm{i}\,\mathrm{m}_{3})^{2}=-(\mathrm{m}_{1})^{2}.
\end{equation}
From Eqs. (29) and (30), we obtain the same results, namely, 
\begin{equation}
\mathrm{m}_{0}+\cos(\mu)=0\quad\mathrm{and}\quad\mathrm{m}_{3}+\sin(\mu)=\pm\,\mathrm{m}_{1}.
\end{equation}
By substituting the latter two equations into Eq. (28), we obtain
the result
\begin{equation}
\mathrm{m}_{3}=0,
\end{equation}
and the condition $(\mathrm{m}_{0})^{2}+(\mathrm{m}_{1})^{2}=1$ is
automatically satisfied, that is, $(-\cos(\mu))^{2}+(\mp\sin(\mu))^{2}=1$
(we recall that $\mathrm{m}_{1}\neq0$; hence, $\mu\neq0$). Finally,
the matrix $\hat{\mathrm{V}}$ in Eq. (26) can be written as follows:
\begin{equation}
\hat{\mathrm{V}}=\mp\frac{\mathrm{i}}{\sin(\mu)}\left[\begin{array}{cc}
-\mathrm{e}^{\mathrm{i}\,\mu} & -\cos(\mu)\\
\cos(\mu) & \mathrm{e}^{-\mathrm{i}\,\mu}
\end{array}\right],
\end{equation}
where $\mu\in(0,\pi)$. The BCs that make up the domain of the free
FV Hamiltonian in Eq. (3), thus far, are those given in Eq. (25),
with the matrix $\hat{\mathrm{V}}$ given in Eq. (33). That is, we
have a one-parameter general family of BCs for this Hamiltonian operator,
namely, $\Phi(b,t)=\hat{\mathrm{V}}\Phi(a,t)$ with $\Phi'(b,t)=\hat{\mathrm{V}}\Phi'(a,t)$.
We recall that this one-parameter family of BCs satisfies the relation
given in Eq. (2), i.e., given a BC within the one-parameter family
of BCs, the energy current density has the same value at each end
of the interval. 

The 1D FV wave equation is given by 
\begin{equation}
\hat{\mathrm{E}}\Phi=\hat{\mathrm{h}}\Phi\quad\Rightarrow\quad\mathrm{i}\hbar\,\hat{1}_{2}\,\frac{\partial}{\partial t}\Phi(x,t)=\left[\,-\frac{\hbar^{2}}{2\mathrm{m}}(\hat{\tau}_{3}+\mathrm{i}\hat{\tau}_{2})\frac{\partial^{2}}{\partial x^{2}}+\mathrm{m}c^{2}\hat{\tau}_{3}\,\right]\Phi(x,t).
\end{equation}
If $\Phi(x,t)$ is a solution of this equation, then 
\begin{equation}
(\hat{\Pi}\Phi)(x,t)=\Phi(a+b-x,t)
\end{equation}
is also a solution. The operator $\hat{\Pi}$ is the parity operator,
and it reflects the wavefunction around the point $x=(a+b)/2$, i.e.,
around the midpoint of the interval. In fact, $(\hat{\Pi}\Phi)(a,t)=\Phi(b,t)$,
$(\hat{\Pi}\Phi)(b,t)=\Phi(a,t)$ and $(\hat{\Pi}\Phi)(\tfrac{a+b}{2},t)=\Phi(\tfrac{a+b}{2},t)$,
as expected. To immediately see that $(\hat{\Pi}\Phi)(x,t)$ is also
a solution of the free FV equation in Eq. (34), the formulas $\tfrac{\partial\Phi(u,t)}{\partial u}=-\tfrac{\partial\Phi(u(x),t)}{\partial x}$,
and therefore, $\tfrac{\partial^{2}\Phi(u,t)}{\partial u^{2}}=+\tfrac{\partial^{2}\Phi(u(x),t)}{\partial x^{2}}$,
where $\Phi$ is a function of $u$ and $t$ and where $u$ is a function
of $x$ (in fact, $u\equiv a+b-x$), can be used. Because $\Phi(x,t)$
and $(\hat{\Pi}\Phi)(x,t)$ are solutions of the free FV equation,
this equation is parity invariant (hence, the Hamiltonian $\hat{\mathrm{h}}$
commutes with $\hat{\Pi}$, i.e., $\hat{\mathrm{h}}(\hat{\Pi}\Phi)=\hat{\Pi}(\hat{\mathrm{h}}\Phi)$
with $\hat{\Pi}\Phi\in\mathrm{\mathcal{D}}(\hat{\mathrm{h}})$). If
$\Phi(x,t)$ and $\Phi'(x,t)$ satisfy a BC, then $(\hat{\Pi}\Phi)(x,t)$
and $(\hat{\Pi}\Phi)'(x,t)$ also satisfy it. Consequently, not only
must the BCs given in Eq. (25) be in $\mathrm{\mathcal{D}}(\hat{\mathrm{h}})$,
but the following must also be included:
\begin{equation}
(\hat{\Pi}\Phi)(b,t)=\hat{\mathrm{V}}\,(\hat{\Pi}\Phi)(a,t)\quad\mathrm{and}\quad(\hat{\Pi}\Phi)'(b,t)=\hat{\mathrm{V}}\,(\hat{\Pi}\Phi)'(a,t),
\end{equation}
where $(\hat{\Pi}\Phi)(x,t)$ at $x=a$ and $x=b$ are obtained from
Eq. (35). Moreover, $(\hat{\Pi}\Phi)'(x,t)=\tfrac{\partial}{\partial x}\left(\Phi(a+b-x,t)\right)=-\left.\tfrac{\partial}{\partial u}\left(\Phi(u,t)\right)\right|_{u=a+b-x}$,
i.e., $(\hat{\Pi}\Phi)'(x,t)=-\Phi'(a+b-x,t)$; hence, $(\hat{\Pi}\Phi)'(a,t)=-\Phi'(b,t)$
and $(\hat{\Pi}\Phi)'(b,t)=-\Phi'(a,t)$. The BCs given in Eq. (36)
can be written as follows:
\begin{equation}
\left[\begin{array}{c}
\phi_{1}(a,t)\\
\phi_{2}(a,t)
\end{array}\right]=\hat{\mathrm{V}}\left[\begin{array}{c}
\phi_{1}(b,t)\\
\phi_{2}(b,t)
\end{array}\right]\quad\mathrm{and}\quad-\left[\begin{array}{c}
\phi_{1}'(a,t)\\
\phi_{2}'(a,t)
\end{array}\right]=-\hat{\mathrm{V}}\left[\begin{array}{c}
\phi_{1}'(b,t)\\
\phi_{2}'(b,t)
\end{array}\right],
\end{equation}
where $\hat{\mathrm{V}}$ is given in Eq. (33). Substituting the latter
two relations into those given in Eq. (25), we obtain a new constraint
on the matrix $\hat{\mathrm{V}}$, namely,
\begin{equation}
\hat{\mathrm{V}}^{2}=\hat{1}_{2}.
\end{equation}
By substituting the matrix $\hat{\mathrm{V}}$ in Eq. (33) into Eq.
(38), the following three equations are obtained:
\begin{equation}
\sin(\mu)\cos(\mu)=0,
\end{equation}
\begin{equation}
\mathrm{e}^{\mathrm{i}\,2\mu}-(\cos(\mu))^{2}=-(\sin(\mu))^{2},
\end{equation}
and
\begin{equation}
\mathrm{e}^{-\mathrm{i}\,2\mu}-(\cos(\mu))^{2}=-(\sin(\mu))^{2},
\end{equation}
and whose only common solution is given by
\begin{equation}
\mu=\frac{\pi}{2}.
\end{equation}
Consequently, the matrix $\hat{\mathrm{V}}$ in Eq. (33) takes the
form
\begin{equation}
\hat{\mathrm{V}}=\mp\,\mathrm{i}\left[\begin{array}{cc}
-\mathrm{i} & 0\\
0 & -\mathrm{i}
\end{array}\right]=\mp\,\hat{1}_{2},
\end{equation}
and the BCs that are part of the domain of the free FV Hamiltonian
are the following:
\begin{equation}
\left[\begin{array}{c}
\phi_{1}(b,t)\\
\phi_{2}(b,t)
\end{array}\right]=\mp\left[\begin{array}{c}
\phi_{1}(a,t)\\
\phi_{2}(a,t)
\end{array}\right]\quad\mathrm{and}\quad\left[\begin{array}{c}
\phi_{1}'(b,t)\\
\phi_{2}'(b,t)
\end{array}\right]=\mp\left[\begin{array}{c}
\phi_{1}'(a,t)\\
\phi_{2}'(a,t)
\end{array}\right],
\end{equation}
that is, the periodic BC, $\Phi(b,t)=\Phi(a,t)$ and $\Phi'(b,t)=\Phi'(a,t)$,
and the antiperiodic BC, $\Phi(b,t)=-\Phi(a,t)$ and $\Phi'(b,t)=-\Phi'(a,t)$.
Owing to the Majorana conditions, it is sufficient to write these
BCs in terms of only one of the components of $\Phi$ (the other component
can be obtained algebraically). For example, in terms of $\phi_{1}$
alone, the periodic BC is $\phi_{1}(b,t)=\phi_{1}(a,t)$ and $\phi_{1}'(b,t)=\phi_{1}'(a,t)$,
and the antiperiodic BC is $\phi_{1}(b,t)=-\phi_{1}(a,t)$ and $\phi_{1}'(b,t)=-\phi_{1}'(a,t)$
(the BCs for $\phi_{2}$ are obtained from the relation $\phi_{2}=\phi_{1}^{*}$,
when $\Phi=\Phi_{c}$, or from $\phi_{2}=-\phi_{1}^{*}$, when $\Phi=-\Phi_{c}$).
Finally, the wave equation for $\phi_{1}$ is obtained from the FV
equation and a Majorana condition, namely, 
\begin{equation}
\mathrm{i}\hbar\,\frac{\partial}{\partial t}\phi_{1}(x,t)=-\frac{\hbar^{2}}{2\mathrm{m}}\frac{\partial^{2}}{\partial x^{2}}\left[\,\phi_{1}(x,t)+\phi_{1}^{*}(x,t)\,\right]+\mathrm{m}c^{2}\,\phi_{1}(x,t),
\end{equation}
when $\Phi=\Phi_{c}$ ($\Rightarrow\phi_{2}=\phi_{1}^{*}$), or 
\begin{equation}
\mathrm{i}\hbar\,\frac{\partial}{\partial t}\phi_{1}(x,t)=-\frac{\hbar^{2}}{2\mathrm{m}}\frac{\partial^{2}}{\partial x^{2}}\left[\,\phi_{1}(x,t)-\phi_{1}^{*}(x,t)\,\right]+\mathrm{m}c^{2}\,\phi_{1}(x,t),
\end{equation}
when $\Phi=-\Phi_{c}$ ($\Rightarrow\phi_{2}=-\phi_{1}^{*}$). Because
the latter two equations describe (strictly) neutral particles, they
do not necessarily lead to the free 1D Schr\"odinger equation in
the nonrelativistic limit, as one might expect. In fact, in this approximation,
Eq. (45) is given by
\begin{equation}
\mathrm{Re}\left[\,\mathrm{e}^{-\mathrm{i}\frac{\mathrm{m}c^{2}}{\hbar}t}\left(\,-\mathrm{i}\hbar\,\frac{\partial}{\partial t}-\frac{\hbar^{2}}{2\mathrm{m}}\frac{\partial^{2}}{\partial x^{2}}\,\right)(\phi_{1})_{\mathrm{NR}}\,\right]=0
\end{equation}
(see Ref. \cite{RefM}). Similarly, the nonrelativistic limit of Eq.
(46) is given by
\begin{equation}
\mathrm{Im}\left[\,\mathrm{e}^{-\mathrm{i}\frac{\mathrm{m}c^{2}}{\hbar}t}\left(\,-\mathrm{i}\hbar\,\frac{\partial}{\partial t}-\frac{\hbar^{2}}{2\mathrm{m}}\frac{\partial^{2}}{\partial x^{2}}\,\right)(\phi_{1})_{\mathrm{NR}}\,\right]=0.
\end{equation}
This result can be obtained via the same procedure that led to Eq.
(47). Certainly, because the terms enclosed in brackets do not have
to be zero, the latter two equations are not free 1D Schr\"odinger
equations. However, if in Eqs. (47) and (48) the functions $(\phi_{1})_{\mathrm{NR}}$
and $(\phi_{1})_{\mathrm{NR}}^{*}$ are interpreted as independent
solutions or fields, then these equations can yield the free Schr\"odinger
equation for $(\phi_{1})_{\mathrm{NR}}$ and the equation that is
the complex conjugate of the free Schr\"odinger equation for $(\phi_{1})_{\mathrm{NR}}^{*}$.
The latter argument has been used and questioned when taking the nonrelativistic
limit of certain real scalar field theories (see, for example, Refs.
\cite{RefT,RefU,RefV,RefW}).

In the free case, the relationship between the two-component FV wavefunction
$\Phi$ and the corresponding one-component KFG wavefunction $\phi$
can be written as follows:
\begin{equation}
\Phi=\left[\begin{array}{c}
\phi_{1}\\
\phi_{2}
\end{array}\right]=\frac{1}{2}\left[\begin{array}{c}
\phi+\frac{1}{\mathrm{m}c^{2}}\hat{\mathrm{E}}\phi\\
\phi-\frac{1}{\mathrm{m}c^{2}}\hat{\mathrm{E}}\phi
\end{array}\right]\quad\Rightarrow\quad(\hat{\tau}_{3}+\mathrm{i}\hat{\tau}_{2})\Phi=\left[\begin{array}{c}
\phi\\
-\phi
\end{array}\right]
\end{equation}
(see, for example, Refs. \cite{RefO,RefP}). The latter matrix relation
precisely leads to the complete equivalence between the KFG and FV
wave equations; certainly, the mass of the particle cannot be equal
to zero. Thus, by virtue of the relation on the right-hand side in
Eq. (49), the two BCs given in Eq. (44), i.e., $(\hat{\tau}_{3}+\mathrm{i}\hat{\tau}_{2})\Phi(b,t)=\mp(\hat{\tau}_{3}+\mathrm{i}\hat{\tau}_{2})\Phi(a,t)$
and $(\hat{\tau}_{3}+\mathrm{i}\hat{\tau}_{2})\Phi'(b,t)=\mp(\hat{\tau}_{3}+\mathrm{i}\hat{\tau}_{2})\Phi'(a,t)$,
lead us to the following two BCs for the one-component wavefunction
$\phi$:
\begin{equation}
\phi(b,t)=\mp\phi(a,t)\quad\mathrm{and}\quad\phi'(b,t)=\mp\phi'(a,t),
\end{equation}
namely, the periodic BC (positive signs) and the antiperiodic BC (negative
signs). Certainly, the KFG one-component wavefunction $\phi$ is real,
or purely imaginary, depending on the Majorana condition chosen. Finally,
because the two BCs in Eq. (50) are included in the most general three-parameter
set of BCs for which the FV Hamiltonian operator for a 1D KFGM particle
in an interval is pseudo self-adjoint, the free FV Hamiltonian operator
for a 1D KFGM particle in an interval with these two BCs is a pseudo
self-adjoint operator (see BCs (vii) and (viii) in Section IV, Ref.
\cite{RefN}).

\section{Final discussion}

\noindent A truly free 1D KFGM particle in an interval is a strictly
neutral 1D particle that is located in a finite region but is not
confined to that region, nor is it subject to any external field.
Consequently, the energy current density $j_{\mathrm{\mathrm{en}}}$
does not vanish at the walls of that region; however, there are infinite
BCs that satisfy this requirement. To distinguish between all these
BCs, we restrict the domain of the free FV Hamiltonian operator, which
is a function of the domain of the pseudo self-adjoint momentum operator
(we calculate the latter domain first) by requiring that, given a
BC within the domain of the Hamiltonian, the energy current density
has the same value at each end of the interval. Then, a one-parameter
general family of BCs must lie within the domain of the free FV Hamiltonian.
Because the FV wave equation is invariant under the operation of parity,
the parity-transformed wavefunction is also a solution of this equation,
which further restricts the domain of the free FV Hamiltonian. Finally,
knowing what BCs are inside the domain of the pseudo self-adjoint
FV Hamiltonian operator for a 1D KFGM particle in an interval, we
find that only two quantum BCs can remain in the domain of the free
FV Hamiltonian: the periodic BC and the antiperiodic BC. These BCs
are satisfied by both the two-component FV wavefunction and the one-component
KFG wavefunction. Furthermore, as a consequence of imposing a Majorana
condition, the two components of the FV wavefunction are not independent
of each other, and the KFG wavefunction is either real or imaginary.
Owing to the Majorana conditions, it is sufficient to write these
BCs in terms of only one of the components of the FV wavefunction.
The first-order equation in time for the chosen component is obtained
from the FV wave equation and a Majorana condition. 

The physical interpretation of the periodic BC imposed on the wavefunctions
$\phi$, $\Phi$, $\phi_{1}$ or $\phi_{2}=\pm\phi_{1}^{*}$, is that
it simulates the motion of a 1D KFGM particle over a complete circle
($360\text{\textdegree}$), i.e., the particle reaches one wall and
reappears at the other wall. In contrast, the physical interpretation
of the antiperiodic BC is that it simulates the motion of the particle
over half of a M\"obius strip ($180\text{\textdegree}$), i.e., the
particle reaches one wall and reappears at the other wall but \textquotedblleft{}upside
down.\textquotedblright{} 

Naturally, when the domain of the momentum operator $\hat{\mathrm{p}}$
contains only the periodic (antiperiodic) BC, i.e., $\Phi(b,t)=\Phi(a,t)$
($\Phi(b,t)=-\Phi(a,t)$), then automatically the domain of the free
FV Hamiltonian operator $\hat{\mathrm{h}}$ that is a function of
$\hat{\mathrm{p}}$ times $\hat{\mathrm{p}}$ contains only the periodic
(antiperiodic) BC, that is, $\Phi(b,t)=\Phi(a,t)$ plus $\Phi'(b,t)=\Phi'(a,t)$
($\Phi(b,t)=-\Phi(a,t)$ plus $\Phi'(b,t)=-\Phi'(a,t)$). Certainly,
both this momentum operator and the free FV Hamiltonian operator are
pseudo self-adjoint operators, and the periodic (antiperiodic) BC
is invariant under the parity operation, i.e., it is a $\hat{\Pi}$-symmetric
BC. These results can be extended to the problem of a truly free 1D
KFGM particle moving in the real line with, for example, the point
$x=0$ excluded. The two BCs for this problem can be obtained from
the two BCs for the truly free 1D KFGM particle in the interval by
making the replacements $x=a\rightarrow0+$ and $x=b\rightarrow0-$.
Certainly, the two nonconfining BCs define transparent or penetrable
barriers at $x=0$. 

Finally, to obtain our results, we use only a few simple concepts
found in the general theory of linear operators over an indefinite
inner product space and, of course, some algebraic procedures to construct
the families and subfamilies of BCs. Furthermore, the main result
of the article, i.e., the result shown in Eq. (44) and in the two
paragraphs following that equation (section III), does not seem to
have been published previously. We believe that our work will be of
interest to all those interested in the fundamental aspects of relativistic
quantum mechanics.

\section*{Appendix A}

\noindent The free 1D Schr\"odinger equation
\[
\mathrm{i}\hbar\,\frac{\partial}{\partial t}\Phi_{\mathrm{S}}(x,t)=\hat{\mathrm{h}}_{\mathrm{S}}\Phi_{\mathrm{S}}(x,t)=\frac{\hat{\mathrm{p}}_{\mathrm{S}}^{2}}{2\mathrm{m}}\Phi_{\mathrm{S}}(x,t)=\frac{\hat{\mathrm{p}}_{\mathrm{S}}\,\hat{\mathrm{p}}_{\mathrm{S}}}{2\mathrm{m}}\Phi_{\mathrm{S}}(x,t)=-\frac{\hbar^{2}}{2\mathrm{m}}\frac{\partial^{2}}{\partial x^{2}}\Phi_{\mathrm{S}}(x,t)\tag{A1}
\]
has as its solution the wavefunction $\Phi_{\mathrm{S}}(x,t)$, but
the parity-transformed wavefunction $(\hat{\Pi}_{\mathrm{S}}\Phi_{\mathrm{S}})(x,t)=\Phi_{\mathrm{S}}(a+b-x,t)$
is also a solution. The procedure used in the paragraph following
Eq. (35) can be repeated here to check the veracity of this result.
The Hamiltonian $\hat{\mathrm{h}}_{\mathrm{S}}$ essentially has the
following domain:
\[
\mathrm{\mathcal{D}}(\hat{\mathrm{h}}_{\mathrm{S}})=\mathrm{\mathcal{D}}(\hat{\mathrm{p}}_{\mathrm{S}}^{2})=\mathrm{\mathcal{D}}(\hat{\mathrm{p}}_{\mathrm{S}}\,\hat{\mathrm{p}}_{\mathrm{S}})\sim\left\{ \,\Phi_{\mathrm{S}}\mid\Phi_{\mathrm{S}}\in\mathrm{\mathcal{D}}(\hat{\mathrm{p}}_{\mathrm{S}})\;\mathrm{and}\;\hat{\mathrm{p}}_{\mathrm{S}}\Phi_{\mathrm{S}}\in\mathrm{\mathcal{D}}(\hat{\mathrm{p}}_{\mathrm{S}})\,\right\} .\tag{A2}
\]
Thus, $\Phi_{\mathrm{S}}\in\mathrm{\mathcal{D}}(\hat{\mathrm{p}}_{\mathrm{S}})$,
therefore, $\Phi_{\mathrm{S}}(b,t)=\mathrm{e}^{\mathrm{i}\,\theta}\,\Phi_{\mathrm{S}}(a,t)$;
and $\hat{\mathrm{p}}_{\mathrm{S}}\Phi_{\mathrm{S}}\in\mathrm{\mathcal{D}}(\hat{\mathrm{p}}_{\mathrm{S}})$,
therefore, $\Phi'_{\mathrm{S}}(b,t)=\mathrm{e}^{\mathrm{i}\,\theta}\,\Phi'_{\mathrm{S}}(a,t)$
(with $\theta\in[0,2\pi)$). Moreover, because $(\hat{\Pi}_{\mathrm{S}}\Phi_{\mathrm{S}})(x,t)$
is also a solution of the equation in (A1), we have the following
restrictions: $\hat{\Pi}_{\mathrm{S}}\Phi_{\mathrm{S}}\in\mathrm{\mathcal{D}}(\hat{\mathrm{p}}_{\mathrm{S}})$,
therefore, $(\hat{\Pi}_{\mathrm{S}}\Phi_{\mathrm{S}})(b,t)=\mathrm{e}^{\mathrm{i}\,\theta}\,(\hat{\Pi}_{\mathrm{S}}\Phi_{\mathrm{S}})(a,t)$;
and $\hat{\mathrm{p}}_{\mathrm{S}}(\hat{\Pi}_{\mathrm{S}}\Phi_{\mathrm{S}})\in\mathrm{\mathcal{D}}(\hat{\mathrm{p}}_{\mathrm{S}})$,
therefore, $(\hat{\Pi}_{\mathrm{S}}\Phi_{\mathrm{S}})'(b,t)=\mathrm{e}^{\mathrm{i}\,\theta}\,(\hat{\Pi}_{\mathrm{S}}\Phi_{\mathrm{S}})'(a,t)$.
The latter two BCs become the following two BCs: $\Phi_{\mathrm{S}}(a,t)=\mathrm{e}^{\mathrm{i}\,\theta}\,\Phi_{\mathrm{S}}(b,t)$
and $\Phi'_{\mathrm{S}}(a,t)=\mathrm{e}^{\mathrm{i}\,\theta}\,\Phi'_{\mathrm{S}}(b,t)$,
respectively. Finally, only two BCs satisfy the two pairs of BCs mentioned
above and therefore only they can be in the domain of the free 1D
Schr\"odinger Hamiltonian: the periodic and antiperiodic BCs, i.e.,
$\Phi_{\mathrm{S}}(b,t)=\Phi_{\mathrm{S}}(a,t)$ plus $\Phi_{\mathrm{S}}'(b,t)=\Phi_{\mathrm{S}}'(a,t)$;
and $\Phi_{\mathrm{S}}(b,t)=-\Phi_{\mathrm{S}}(a,t)$ plus $\Phi_{\mathrm{S}}'(b,t)=-\Phi_{\mathrm{S}}'(a,t)$,
respectively.

\section*{Appendix B}

\noindent The FV Hamiltonian operator for a 1D KFG particle in the
interval $\Omega=[a,b]$ is given by 
\[
\hat{\mathrm{h}}=\frac{\hat{\mathrm{p}}^{2}}{2\mathrm{m}}(\hat{\tau}_{3}+\mathrm{i}\hat{\tau}_{2})+\mathrm{m}c^{2}\hat{\tau}_{3}.\tag{B1}
\]
The formal generalized adjoint of $\hat{\mathrm{h}}$ is given by
\[
\hat{\mathrm{h}}_{\mathrm{adj}}\equiv\hat{\tau}_{3}\,\hat{\mathrm{h}}^{\dagger}\,\hat{\tau}_{3}=\frac{\hat{\mathrm{p}}^{2}}{2\mathrm{m}}(\hat{\tau}_{3}+\mathrm{i}\hat{\tau}_{2})+\mathrm{m}c^{2}\hat{\tau}_{3}.\tag{B2}
\]
Clearly, the actions of $\hat{\mathrm{h}}$ and $\hat{\mathrm{h}}_{\mathrm{adj}}$
are equal. The latter definition leads automatically to the following
relation:
\[
\langle\langle\hat{\mathrm{h}}_{\mathrm{adj}}\Psi,\Phi\rangle\rangle=\langle\langle\Psi,\hat{\mathrm{h}}\Phi\rangle\rangle,\tag{B3}
\]
where the scalar product is defined in Eq. (7). The latter relation
defines the operator $\hat{\mathrm{h}}_{\mathrm{adj}}$ on a $\hat{\tau}_{3}$-space,
where $\Psi\in\mathrm{\mathcal{D}}(\hat{\mathrm{h}}_{\mathrm{adj}})$,
the domain of $\hat{\mathrm{h}}_{\mathrm{adj}}$, and $\Phi\in\mathrm{\mathcal{D}}(\hat{\mathrm{h}})$,
the domain of $\hat{\mathrm{h}}$. By applying the method of integration
by parts twice, the following relation can be demonstrated \cite{RefX}:
\[
\langle\langle\hat{\mathrm{h}}_{\mathrm{adj}}\Psi,\Phi\rangle\rangle=\langle\langle\Psi,\hat{\mathrm{h}}\Phi\rangle\rangle+g[\Psi,\Phi;\Omega],\tag{B4}
\]
where the boundary term is given by 
\[
g[\Psi,\Phi;\Omega]=-\frac{\hbar^{2}}{2\mathrm{m}}\,\frac{1}{2}\left.\left[\,\left((\hat{\tau}_{3}+\mathrm{i}\hat{\tau}_{2})\Psi'\right)^{\dagger}(\hat{\tau}_{3}+\mathrm{i}\hat{\tau}_{2})\Phi-\left((\hat{\tau}_{3}+\mathrm{i}\hat{\tau}_{2})\Psi\right)^{\dagger}(\hat{\tau}_{3}+\mathrm{i}\hat{\tau}_{2})\Phi'\,\right]\right|_{a}^{b}.\tag{B5}
\]
Using the relationship between the two-component FV wavefunction $\Phi$
and the corresponding one-component KFG wavefunction $\phi$ given
in Eq. (49) and another similar relationship for $\Psi$ and $\psi$,
we can write $g[\Psi,\Phi;\Omega]$ in terms of only $\psi$ and $\phi$,
namely, 
\[
g[\Psi,\Phi;\Omega]=-\frac{\hbar^{2}}{2\mathrm{m}}\left.\left[\,(\psi')^{*}\,\phi-\psi^{*}\phi'\,\right]\right|_{a}^{b}.\tag{B6}
\]

The most general family of BCs for a 1D KFG particle in an interval,
characterized by four real parameters and leading to the disappearance
of $g[\Psi,\Phi;\Omega]$, was obtained in Ref. \cite{RefX}, namely,
\[
\left[\begin{array}{c}
\phi(b,t)-\mathrm{i}\lambda\phi'(b,t)\\
\phi(a,t)+\mathrm{i}\lambda\phi'(a,t)
\end{array}\right]=\hat{\mathrm{U}}\left[\begin{array}{c}
\phi(b,t)+\mathrm{i}\lambda\phi'(b,t)\\
\phi(a,t)-\mathrm{i}\lambda\phi'(a,t)
\end{array}\right],\tag{B7}
\]
where $\phi=\phi_{1}+\phi_{2}$ (see Eq. (49)), $\lambda$ is a real
parameter and the unitary matrix $\hat{\mathrm{U}}$ is given by 
\[
\hat{\mathrm{U}}=\mathrm{e}^{\mathrm{i}\,\theta}\left[\begin{array}{cc}
\mathrm{n}_{0}-\mathrm{i}\,\mathrm{n}_{3} & -\mathrm{n}_{2}-\mathrm{i}\,\mathrm{n}_{1}\\
\mathrm{n}_{2}-\mathrm{i}\,\mathrm{n}_{1} & \mathrm{n}_{0}+\mathrm{i}\,\mathrm{n}_{3}
\end{array}\right],\tag{B8}
\]
where $\theta\in[0,\pi)$, and the real parameters $\mathrm{n}_{0}$,
$\mathrm{n}_{1}$, $\mathrm{n}_{2}$ and $\mathrm{n}_{3}$ satisfy
$(\mathrm{n}_{0})^{2}+(\mathrm{n}_{1})^{2}+(\mathrm{n}_{2})^{2}+(\mathrm{n}_{3})^{2}=1$.
By imposing a Majorana condition on the general family of BCs given
in Eqs. (B7) and (B8) ($\Phi=\Phi_{c}=\hat{\tau}_{1}\Phi^{*}$ $\Rightarrow\phi=\phi^{*}$
or $\Phi=-\Phi_{c}=-\hat{\tau}_{1}\Phi^{*}$ $\Rightarrow\phi=-\phi^{*}$),
the result $\mathrm{n}_{2}=0$ is obtained. Thus, the four-parameter
family of BCs becomes a three-parameter family of BCs. Indeed, the
latter family is for a 1D KFGM particle in an interval \cite{RefM}.
For all BCs included in the general set of BCs given in Eqs. (B7)
and (B8), $\hat{\mathrm{h}}$ is a pseudo self-adjoint operator, i.e.,
\[
\langle\langle\hat{\mathrm{h}}\Psi,\Phi\rangle\rangle=\langle\langle\Psi,\hat{\mathrm{h}}\Phi\rangle\rangle.\tag{B9}
\]
Thus, $\hat{\mathrm{h}}=\hat{\mathrm{h}}_{\mathrm{adj}}$, i.e., the
actions of $\hat{\mathrm{h}}$ and $\hat{\mathrm{h}}_{\mathrm{adj}}$
are equal, and $\mathrm{\mathcal{D}}(\hat{\mathrm{h}})=\mathrm{\mathcal{D}}(\hat{\mathrm{h}}_{\mathrm{adj}})$. 

If we make $\Psi=\Phi$ in Eq. (B9) and in Eq. (B4) with the result
$\hat{\mathrm{h}}=\hat{\mathrm{h}}_{\mathrm{adj}}$, we obtain the
following condition \cite{RefX}:
\[
g[\Phi,\Phi;\Omega]=\mathrm{i}\hbar\left.\left[\, j\,\right]\right|_{a}^{b}=0\quad\Rightarrow\quad j(b,t)=j(a,t)\,,\tag{B10}
\]
where the usual KFG current density is given by
\[
j=j(x,t)=\frac{\mathrm{i}\hbar}{2\mathrm{m}}\,\frac{1}{2}\left[\,\left((\hat{\tau}_{3}+\mathrm{i}\hat{\tau}_{2})\Phi'\right)^{\dagger}(\hat{\tau}_{3}+\mathrm{i}\hat{\tau}_{2})\Phi-\left((\hat{\tau}_{3}+\mathrm{i}\hat{\tau}_{2})\Phi\right)^{\dagger}(\hat{\tau}_{3}+\mathrm{i}\hat{\tau}_{2})\Phi'\,\right].\tag{B11}
\]
The operators $\hat{\mathrm{h}}$ and $\hat{\mathrm{h}}_{\mathrm{adj}}$
also satisfy the following relation:
\[
\langle\langle\hat{\mathrm{h}}_{\mathrm{adj}}\Psi,\hat{\mathrm{E}}\Phi\rangle\rangle=\langle\langle\Psi,\hat{\mathrm{h}}\,\hat{\mathrm{E}}\Phi\rangle\rangle+g[\Psi,\hat{\mathrm{E}}\Phi;\Omega],\tag{B12}
\]
where $\hat{\mathrm{E}}=\mathrm{i}\hbar\,\hat{1}_{2}\,\partial/\partial t$
is the energy operator. If $\Phi\in\mathrm{\mathcal{D}}(\hat{\mathrm{h}})$,
then automatically, $\hat{\mathrm{E}}\Phi\in\mathrm{\mathcal{D}}(\hat{\mathrm{h}})$.
Thus, any BC within the general set of pseudo self-adjoint BCs given
in Eqs. (B7) and (B8) that leads to the disappearance of the boundary
term $g[\Psi,\Phi;\Omega]$ in Eq. (B4) must also lead to the disappearance
of $g[\Psi,\hat{\mathrm{E}}\Phi;\Omega]$ in Eq. (B12). If we make
$\Psi=\Phi$ in Eq. (B12) with the result $\hat{\mathrm{h}}=\hat{\mathrm{h}}_{\mathrm{adj}}$,
we obtain the following result \cite{RefN}:
\[
g[\Phi,\hat{\mathrm{E}}\Phi;\Omega]=\mathrm{i}\hbar\left.\left[\, j_{\mathrm{en}}\,\right]\right|_{a}^{b}=0\quad\Rightarrow\quad j_{\mathrm{en}}(b,t)=j_{\mathrm{en}}(a,t)\,,\tag{B13}
\]
where the (complex) KFG energy current density is precisely the result
given in Eq. (1), namely, 
\[
j_{\mathrm{en}}=j_{\mathrm{en}}(x,t)=-\frac{\hbar^{2}}{2\mathrm{m}}\,\frac{1}{2}\left[\,\left((\hat{\tau}_{3}+\mathrm{i}\hat{\tau}_{2})\Phi'\right)^{\dagger}(\hat{\tau}_{3}+\mathrm{i}\hat{\tau}_{2})\dot{\Phi}-\left((\hat{\tau}_{3}+\mathrm{i}\hat{\tau}_{2})\Phi\right)^{\dagger}(\hat{\tau}_{3}+\mathrm{i}\hat{\tau}_{2})\dot{\Phi}'\,\right].\tag{B14}
\]
The results given in Eqs. (B10) and (B13) are direct consequences
of the pseudo self-adjointness of $\hat{\mathrm{h}}$. Moreover, $j$
is equal to zero everywhere and $j_{\mathrm{en}}$ is a real quantity
when $j$ and $j_{\mathrm{en}}$ are calculated for a state $\Phi$
satisfying a Majorana condition. 

The following relations are also obtained:
\[
\frac{\mathrm{d}}{\mathrm{d}t}\langle\langle\Psi,\Phi\rangle\rangle=-\frac{1}{\mathrm{i}\hbar}\, g[\Psi,\Phi;\Omega]\tag{B15}
\]
and 
\[
\frac{\mathrm{d}}{\mathrm{d}t}\langle\langle\Psi,\hat{\mathrm{E}}\Phi\rangle\rangle=-\frac{1}{\mathrm{i}\hbar}\, g[\Psi,\hat{\mathrm{E}}\Phi;\Omega],\tag{B16}
\]
where $\Psi$ and $\Phi$ in Eqs. (B15) and (B16) must also satisfy
the 1D FV wave equation. If $\hat{\mathrm{h}}$ is a pseudo self-adjoint
operator, then the boundary terms in Eqs. (B15) and (B16) vanish;
consequently, $\langle\langle\Psi,\Phi\rangle\rangle$ and $\langle\langle\Psi,\hat{\mathrm{E}}\Phi\rangle\rangle$
are time-independent quantities. Moreover, if we make $\Psi=\Phi$
in Eqs. (B15) and (B16), the following results are obtained:
\[
\frac{\mathrm{d}}{\mathrm{d}t}\langle\langle\Phi,\Phi\rangle\rangle=\frac{\mathrm{d}}{\mathrm{d}t}\left(\int_{\Omega}\mathrm{d}x\,\varrho\right)=-\left.\left[\, j\,\right]\right|_{a}^{b}\tag{B17}
\]
and 
\[
\frac{\mathrm{d}}{\mathrm{d}t}\langle\langle\Phi,\hat{\mathrm{E}}\Phi\rangle\rangle=\frac{\mathrm{d}}{\mathrm{d}t}\left(\int_{\Omega}\mathrm{d}x\,\varrho_{\mathrm{E}}\right)=-\left.\left[\, j_{\mathrm{E}}\,\right]\right|_{a}^{b},\tag{B18}
\]
where the indefinite scalar product definition in Eq. (7) was also
used. If $\hat{\mathrm{h}}$ is a pseudo self-adjoint operator, i.e.,
if the results in Eqs. (B10) and (B13) are imposed, then $\langle\langle\Phi,\Phi\rangle\rangle=\int_{\Omega}\mathrm{d}x\,\varrho=\mathrm{const}$
and $\langle\langle\Phi,\hat{\mathrm{E}}\Phi\rangle\rangle=\int_{\Omega}\mathrm{d}x\,\varrho_{\mathrm{E}}=\mathrm{const}$
(on-shell results). In fact, the results in Eqs. (B17) and (B18) are
obtained by integrating the continuity equations $\hat{\mathrm{E}}\varrho-\hat{\mathrm{p}}j=0$
and $\hat{\mathrm{E}}\varrho_{\mathrm{E}}-\hat{\mathrm{p}}j_{\mathrm{E}}=0$,
respectively, over the interval $\Omega$.

The usual Lorentz two-vector current density is given by
\[
j^{\mu}=j^{\mu}(x,t)=\frac{\mathrm{i}\hbar}{2\mathrm{m}}[\,\phi^{*}(\partial^{\mu}\phi)-(\partial^{\mu}\phi^{*})\,\phi\,],\tag{B19}
\]
where $\phi=\phi(x,t)$ is a one-component KFG wavefunction, $j^{0}=c\varrho$
and $j^{1}=j$. In (1+1) dimensions, Greek indices are restricted
to $\mu,\nu,\ldots,\mathrm{\{etc.\}}=0,1.$, and the metric tensor
is $g^{\mu\nu}=\mathrm{diag}(1,-1)$. The continuity equation then
takes the usual form $\partial_{\mu}j^{\mu}=0$ \cite{RefO}; therefore,
$\hat{\mathrm{E}}\varrho-\hat{\mathrm{p}}j=0$ (where $\hat{\mathrm{E}}=\mathrm{i}\hbar c\,\partial_{0}$
and $\hat{\mathrm{p}}=-\mathrm{i}\hbar\,\partial_{1}$). We introduce
the following second-rank Lorentz tensor:
\[
K_{\;\;\nu}^{\mu}=K_{\;\;\nu}^{\mu}(x,t)=\frac{\mathrm{i}\hbar}{2\mathrm{m}}\left[\,\phi^{*}\left(\partial^{\mu}(\mathrm{i}\hbar\,\partial_{\nu}\phi)\right)-(\partial^{\mu}\phi^{*})\,(\mathrm{i}\hbar\,\partial_{\nu}\phi)\,\right],\tag{B20}
\]
where $K_{\;\;0}^{0}=\varrho_{\mathrm{E}}$ and $K_{\;\;0}^{1}=j_{\mathrm{E}}/c$;
also, $K^{01}=K^{10}$ when $\psi$ satisfies a Majorana condition
($\phi=\phi^{*}$, or $\phi=-\phi^{*}$), i.e., in this case, $K^{\mu\nu}$
is a symmetric tensor. The latter tensor is an unusual energy-momentum
tensor. However, $K_{\;\;\nu}^{\mu}$ and the commonly used energy-momentum
tensor \cite{RefY}
\[
T_{\;\;\nu}^{\mu}=T_{\;\;\nu}^{\mu}(x,t)=\frac{\hbar^{2}}{2\mathrm{m}}\left[\,(\partial_{\nu}\phi^{*})\,(\partial^{\mu}\phi)+(\partial^{\mu}\phi^{*})\,(\partial_{\nu}\phi)\,\right]-\frac{\hbar^{2}}{2\mathrm{m}}\left[\,(\partial^{\alpha}\phi^{*})\,(\partial_{\alpha}\phi)-\frac{\mathrm{m}^{2}c^{2}}{\hbar^{2}}\phi^{*}\phi\,\right]g_{\;\;\nu}^{\mu}\tag{B21}
\]
are linked via the following relation: 
\[
K_{\;\;\nu}^{\mu}=T_{\;\;\nu}^{\mu}+\frac{\hbar^{2}}{2\mathrm{m}}\left[\,(\partial^{\alpha}\phi^{*})\,(\partial_{\alpha}\phi)-\frac{\mathrm{m}^{2}c^{2}}{\hbar^{2}}\phi^{*}\phi\,\right]g_{\;\;\nu}^{\mu}-\frac{\hbar^{2}}{2\mathrm{m}}\,\partial_{\nu}\left(\phi^{*}(\partial^{\mu}\phi)\right).\tag{B22}
\]
Using the free KFG equation, 
\[
\left(\,\partial^{\alpha}\partial_{\alpha}+\frac{\mathrm{m}^{2}c^{2}}{\hbar^{2}}\,\right)\phi=0,\tag{B23}
\]
the following result can be obtained from Eq. (B22):
\[
\partial_{\mu}K_{\;\;\nu}^{\mu}=\partial_{\mu}T_{\;\;\nu}^{\mu}.\tag{B24}
\]
As is well known, $\partial_{\mu}T_{\;\;\nu}^{\mu}=0$ \cite{RefY},
consequently, 
\[
\partial_{\mu}K_{\;\;\nu}^{\mu}=0.\tag{B25}
\]
By choosing $\nu=0$, the latter equation gives us a specific continuity
equation, i.e., $\partial_{0}K_{\;\;0}^{0}+\partial_{1}K_{\;\;0}^{1}=0$;
therefore, $\hat{\mathrm{E}}\varrho_{\mathrm{E}}-\hat{\mathrm{p}}j_{\mathrm{E}}=0$,
as expected. 

\section*{Appendix C}

\noindent The BCs corresponding to the case $\mathrm{m}_{1}=0$ can
be obtained from the general family of BCs given in Eq. (24) with
the matrix $\hat{\mathrm{N}}$ given in Eq. (23). After setting $\mathrm{m}_{1}=0$,
the following two-parameter family of BCs is obtained:
\[
\hat{\mathrm{V}}_{1}\left[\begin{array}{c}
\phi_{1}(b,t)\\
\phi_{2}(b,t)
\end{array}\right]=\hat{\mathrm{V}}_{2}\left[\begin{array}{c}
\phi_{1}(a,t)\\
\phi_{2}(a,t)
\end{array}\right]\quad\mathrm{and}\quad\hat{\mathrm{V}}_{1}\left[\begin{array}{c}
\phi_{1}'(b,t)\\
\phi_{2}'(b,t)
\end{array}\right]=\hat{\mathrm{V}}_{2}\left[\begin{array}{c}
\phi_{1}'(a,t)\\
\phi_{2}'(a,t)
\end{array}\right],\tag{C1}
\]
where 
\[
\hat{\mathrm{V}}_{1}=\left[\begin{array}{cc}
1 & \:-\mathrm{e}^{\mathrm{i}\,\mu}(\mathrm{m}_{0}-\mathrm{i}\,\mathrm{m}_{3})\\
0 & \:0
\end{array}\right]\quad\mathrm{and}\quad\hat{\mathrm{V}}_{2}=\left[\begin{array}{cc}
0 & \:0\\
1 & \:-\mathrm{e}^{-\mathrm{i}\,\mu}(\mathrm{m}_{0}-\mathrm{i}\,\mathrm{m}_{3})
\end{array}\right],\tag{C2}
\]
where $(\mathrm{m}_{0})^{2}+(\mathrm{m}_{3})^{2}=1$, and $\mu\in[0,\pi)$
($\mathrm{det}(\hat{\mathrm{V}}_{1})=\mathrm{det}(\hat{\mathrm{V}}_{2})=0$).
The latter set of BCs may be called ``separated'' because each of
the relations arising from Eq. (C1) is always evaluated at a single
point, that is, at $x=a$ or at $x=b$.

Then, evaluating the energy current density given in Eq. (1) at $x=b$,
substituting into there the relations $\phi_{1}(b,t)=\mathrm{e}^{\mathrm{i}\,\mu}(\mathrm{m}_{0}-\mathrm{i}\,\mathrm{m}_{3})\phi_{2}(b,t)$
and $\phi_{1}'(b,t)=\mathrm{e}^{\mathrm{i}\,\mu}(\mathrm{m}_{0}-\mathrm{i}\,\mathrm{m}_{3})\phi_{2}'(b,t)$
(obtained from the two relations in Eq. (C1)) and their time derivatives,
we obtain the result
\[
j_{\mathrm{en}}(b,t)=-\frac{\hbar^{2}}{\mathrm{m}}\left[\,1+\mathrm{m}_{0}\cos(\mu)+\mathrm{m}_{3}\sin(\mu)\,\right]\left[(\phi_{2}')^{*}(b,t)\,\dot{\phi}_{2}(b,t)-(\phi_{2})^{*}(b,t)\,\dot{\phi}_{2}'(b,t)\right].\tag{C3}
\]
Similarly, evaluating the energy current density given in Eq. (1)
at $x=a$, substituting into there the relations $\phi_{1}(a,t)=\mathrm{e}^{-\mathrm{i}\,\mu}(\mathrm{m}_{0}-\mathrm{i}\,\mathrm{m}_{3})\phi_{2}(a,t)$
and $\phi_{1}'(a,t)=\mathrm{e}^{-\mathrm{i}\,\mu}(\mathrm{m}_{0}-\mathrm{i}\,\mathrm{m}_{3})\phi_{2}'(a,t)$
(obtained from the two relations in Eq. (C1)) and their time derivatives,
we obtain the result
\[
j_{\mathrm{en}}(a,t)=-\frac{\hbar^{2}}{\mathrm{m}}\left[\,1+\mathrm{m}_{0}\cos(\mu)-\mathrm{m}_{3}\sin(\mu)\,\right]\left[(\phi_{2}')^{*}(a,t)\,\dot{\phi}_{2}(a,t)-(\phi_{2})^{*}(a,t)\,\dot{\phi}_{2}'(a,t)\right].\tag{C4}
\]
Because the latter two results must be equal (see Eq. (2)), we have
that $\mathrm{m}_{3}\sin(\mu)=0$; hence, 
\[
\mathrm{m}_{3}=0\:,\quad\mu=0\quad\mathrm{and}\quad\mathrm{m}_{0}=\pm1.\tag{C5}
\]
Therefore, the BCs that emerge from the latter restrictions are given
by
\[
\left[\begin{array}{cc}
1 & \mp1\\
0 & 0
\end{array}\right]\Phi(b,t)=\left[\begin{array}{cc}
0 & 0\\
1 & \mp1
\end{array}\right]\Phi(a,t)\quad\mathrm{and}\quad\left[\begin{array}{cc}
1 & \mp1\\
0 & 0
\end{array}\right]\Phi'(b,t)=\left[\begin{array}{cc}
0 & 0\\
1 & \mp1
\end{array}\right]\Phi'(a,t),\tag{C6}
\]
where the upper (lower) sign corresponds to $\mathrm{m}_{0}=+1$ ($\mathrm{m}_{0}=-1$).
Note that if $\mathrm{m}_{0}=+1$, then the two components of $\Phi(x,t)$,
$\phi_{1}(x,t)$ and $\phi_{2}(x,t)$, are equal at $x=b$, and their
first derivatives, $\phi_{1}'(x,t)$ and $\phi_{2}'(x,t)$, are also
equal there; the same happens at $x=a$. If $\mathrm{m}_{0}=-1$,
then the components of $\Phi(x,t)$ and their first spatial derivatives
have opposite signs at $x=b$, and the same occurs at $x=a$. The
BCs in Eq. (C6) can be written in terms of the one-component KFG wavefunction
$\phi$ (indeed, we use the equivalence between the KFG and FV wave
equations). Evaluating the wavefunction $\Phi$ at $x=b$ and $x=a$
in the relation on the left side of Eq. (47) and substituting these
quantities into Eq. (C6), we obtain the following two BCs:
\[
\dot{\phi}(b,t)=\dot{\phi}(a,t)=0\quad\mathrm{and}\quad\dot{\phi}'(b,t)=\dot{\phi}'(a,t)=0\tag{C7}
\]
($\mathrm{m}_{0}=+1$). This BC is the time derivative of the Dirichlet
and Neumann BCs, and 
\[
\phi(b,t)=\phi(a,t)=0\quad\mathrm{and}\quad\phi'(b,t)=\phi'(a,t)=0\tag{C8}
\]
($\mathrm{m}_{0}=-1$). This BC is the Dirichlet and Neumann BCs. 

The energy current density given in Eq. (1) can also be written in
terms of the wavefunction $\phi$ \cite{RefN}, namely,
\[
j_{\mathrm{en}}=j_{\mathrm{en}}(x,t)=-\frac{\hbar^{2}}{2\mathrm{m}}\left[\,(\phi')^{*}\,\dot{\phi}-\phi^{*}\,\dot{\phi}'\,\right].\tag{C9}
\]
Clearly, if $j_{\mathrm{en}}$ is evaluated at the walls of the interval
and the BC given in Eq. (C7) is used, then the impenetrability condition
is obtained, i.e., $j_{\mathrm{en}}(b,t)=j_{\mathrm{en}}(a,t)=0$.
The same applies if the BC given in Eq. (C8) is used. Therefore, these
two BCs do not characterize a truly free 1D KFGM particle moving in
an interval. Thus, the case $\mathrm{m}_{1}=0$ does not generate
new nonconfining BCs. However, these two BCs, i.e., the BC in Eq.
(C7) (or the BC in Eq. (C6) with the upper sign) and the BC in Eq.
(C8) (or the BC in Eq. (C6) with the lower sign), are not included
in the most general three-parameter family of BCs for the pseudo self-adjoint
FV Hamiltonian operator that describes a 1D KFGM particle in an interval
(see Section IV in Ref. \cite{RefN}). Consequently, these two quantum
BCs cannot be part of the domain of the pseudo self-adjoint--free
FV Hamiltonian operator and must be discarded.

\section*{Conflicts of interest}

\noindent The authors declare no conflicts of interest.

\section*{Data Availability Statement}

\noindent No datasets were generated or analyzed during the current
study.

\section*{Authors' Contributions}

\noindent Conceptualization, T.K. and S.DeV.; methodology, S.DeV.;
formal analysis, T.K. and S.DeV.; investigation, T.K. and S.DeV.;
resources, T.K. and S.DeV.; writing--original draft preparation, S.DeV.;
writing--review and editing, T.K. and S.DeV.; supervision, S.DeV.
Both authors have read and agreed to the current version of the manuscript.

\end{document}